\documentclass[english,reprint,twocolumn,prl, superscriptaddress]{revtex4-1}
\usepackage[T1]{fontenc}
\usepackage[latin9]{inputenc}
\usepackage{amsmath}
\usepackage{amsthm}

\makeatletter

\usepackage{braket}
\usepackage{bbm}

\usepackage{bm}
\usepackage[unicode=true, breaklinks=false, pdfborder={0 0 1}, backref=false, colorlinks=true, linkcolor=blue, urlcolor=blue, citecolor=blue]{hyperref}

\makeatother

\usepackage{babel}
\begin{document}

\title{Reply to Comment on ``Strong Quantum Darwinism and Strong Independence are Equivalent to Spectrum Broadcast Structure''}

\author{Thao P. Le}
\affiliation{Department of Physics and Astronomy, University College London, Gower Street, London WC1E 6BT, United Kingdom}
\affiliation{School of Mathematical Sciences, University of Nottingham, University Park Campus, Nottingham NG7 2RD, United Kingdom}

\author{Alexandra Olaya-Castro}
\affiliation{Department of Physics and Astronomy, University College London, Gower Street, London WC1E 6BT, United Kingdom}

\maketitle

In a recent comment \citep{Feller2021}  on our Letter \citep{Le2019}, \citet{Feller2021} identified a mistake in our mathematical expression of ``strong independence'' for system-environment states that satisfy Spectrum Broadcast Structure. We concede that we wrote a mathematical condition that is necessary but not sufficient. However, we used the original and correct qualitative definition for ``strong independence'' throughout the paper and in our proofs, therefore the proofs and statements, aside from the aforementioned mathematical expression, remain correct.

The concept of ``strong independence'' was originally, qualitatively, introduced by \citet{Horodecki2015} as the situation where ``\emph{...the only correlation between the environments should be the common information about the system. In other words, conditioned by the information about the system, there should be no correlations between the environments.}''
We adhered to this qualitative definition all throughout our paper and our proofs of the connection between Strong Quantum Darwinism and Spectrum Broadcast Structure  \citep{Le2019} and, therefore, our proofs in \citep{Le2019} contain no error. 

The mistake in our paper is in the formal expression Eq.~(9) in \citep{Le2019} for the written statement of strong independence. 
As the authors of the comment \citep{Feller2021} indicate, the correct formal expression for Eq.~(9) should be given by the conditional multipartite mutual information, rather than the conditional pairwise mutual information, for when there are more than two environments. Therefore, Eq.~(9) in \citep{Le2019} shall read
$$I\mathcal{(E}_{1},...,\mathcal{E}_F|\mathcal{S})=0,$$  
which is consistent with the statement we wrote in our paper right after Eq.~(9): 
``\emph{Strong independence means that there are no correlations between the environments conditioned on the information about the system.}''

We also note that the purpose of our original paper Ref. \citep{Le2019} was twofold: to introduce the concept of \emph{Strong Quantum Darwinism} and to make the connection with  Spectrum Broadcast Structure \citep{Horodecki2015}, i.e., Strong Quantum Darwinism with strong independence is equivalent to Spectrum Broadcast Structure. Since Strong Quantum Darwinism is not at all conditioned on strong independence, our proposed Strong Quantum Darwinism framework remains completely untouched by the comment \citep{Feller2021}, both conceptually and mathematically. Its importance in establishing the minimum conditions for the objectivity of a state therefore remains true.

In conclusion, we are grateful to the authors of the comment \citep{Feller2021} for catching the error in Eq.~(9) in the original paper \citep{Le2019} and replacing it with the correct mathematical expression.

We thank the authors for communicating with us ahead of the publication of their comment. T. P. L. acknowledges support from the Engineering and Physical Sciences Research Council. A. O.-C. acknowledges support from the Gordon and Betty Moore Foundation [Grant number GBMF8820].

\bibliographystyle{apsrev4-1}
\bibliography{biblio}

\begin{thebibliography}{3}%
\makeatletter
\providecommand \@ifxundefined [1]{%
 \@ifx{#1\undefined}
}%
\providecommand \@ifnum [1]{%
 \ifnum #1\expandafter \@firstoftwo
 \else \expandafter \@secondoftwo
 \fi
}%
\providecommand \@ifx [1]{%
 \ifx #1\expandafter \@firstoftwo
 \else \expandafter \@secondoftwo
 \fi
}%
\providecommand \natexlab [1]{#1}%
\providecommand \enquote  [1]{``#1''}%
\providecommand \bibnamefont  [1]{#1}%
\providecommand \bibfnamefont [1]{#1}%
\providecommand \citenamefont [1]{#1}%
\providecommand \href@noop [0]{\@secondoftwo}%
\providecommand \href [0]{\begingroup \@sanitize@url \@href}%
\providecommand \@href[1]{\@@startlink{#1}\@@href}%
\providecommand \@@href[1]{\endgroup#1\@@endlink}%
\providecommand \@sanitize@url [0]{\catcode `\\12\catcode `\$12\catcode
  `\&12\catcode `\#12\catcode `\^12\catcode `\_12\catcode `\%12\relax}%
\providecommand \@@startlink[1]{}%
\providecommand \@@endlink[0]{}%
\providecommand \url  [0]{\begingroup\@sanitize@url \@url }%
\providecommand \@url [1]{\endgroup\@href {#1}{\urlprefix }}%
\providecommand \urlprefix  [0]{URL }%
\providecommand \Eprint [0]{\href }%
\providecommand \doibase [0]{http://dx.doi.org/}%
\providecommand \selectlanguage [0]{\@gobble}%
\providecommand \bibinfo  [0]{\@secondoftwo}%
\providecommand \bibfield  [0]{\@secondoftwo}%
\providecommand \translation [1]{[#1]}%
\providecommand \BibitemOpen [0]{}%
\providecommand \bibitemStop [0]{}%
\providecommand \bibitemNoStop [0]{.\EOS\space}%
\providecommand \EOS [0]{\spacefactor3000\relax}%
\providecommand \BibitemShut  [1]{\csname bibitem#1\endcsname}%
\let\auto@bib@innerbib\@empty
\bibitem [{\citenamefont {Feller}\ \emph {et~al.}(2021)\citenamefont {Feller},
  \citenamefont {Roussel}, \citenamefont {Fr\'{e}rot},\ and\ \citenamefont
  {Degiovanni}}]{Feller2021}%
  \BibitemOpen
  \bibfield  {author} {\bibinfo {author} {\bibfnamefont {A.}~\bibnamefont
  {Feller}}, \bibinfo {author} {\bibfnamefont {B.}~\bibnamefont {Roussel}},
  \bibinfo {author} {\bibfnamefont {I.}~\bibnamefont {Fr\'{e}rot}}, \ and\
  \bibinfo {author} {\bibfnamefont {P.}~\bibnamefont {Degiovanni}},\
  }\href@noop {} {\  (\bibinfo {year} {2021})},\ \Eprint
  {http://arxiv.org/abs/arXiv:2101.09186v1} {arXiv:2101.09186v1} \BibitemShut
  {NoStop}%
\bibitem [{\citenamefont {Le}\ and\ \citenamefont
  {Olaya-Castro}(2019)}]{Le2019}%
  \BibitemOpen
  \bibfield  {author} {\bibinfo {author} {\bibfnamefont {T.~P.}\ \bibnamefont
  {Le}}\ and\ \bibinfo {author} {\bibfnamefont {A.}~\bibnamefont
  {Olaya-Castro}},\ }\href {\doibase 10.1103/physrevlett.122.010403} {\bibfield
   {journal} {\bibinfo  {journal} {Phys. Rev. Lett.}\ }\textbf {\bibinfo
  {volume} {122}},\ \bibinfo {pages} {010403} (\bibinfo {year}
  {2019})}\BibitemShut {NoStop}%
\bibitem [{\citenamefont {Horodecki}\ \emph {et~al.}(2015)\citenamefont
  {Horodecki}, \citenamefont {Korbicz},\ and\ \citenamefont
  {Horodecki}}]{Horodecki2015}%
  \BibitemOpen
  \bibfield  {author} {\bibinfo {author} {\bibfnamefont {R.}~\bibnamefont
  {Horodecki}}, \bibinfo {author} {\bibfnamefont {J.~K.}\ \bibnamefont
  {Korbicz}}, \ and\ \bibinfo {author} {\bibfnamefont {P.}~\bibnamefont
  {Horodecki}},\ }\href {\doibase 10.1103/physreva.91.032122} {\bibfield
  {journal} {\bibinfo  {journal} {Phys. Rev. A}\ }\textbf {\bibinfo {volume}
  {91}},\ \bibinfo {pages} {032122} (\bibinfo {year} {2015})}\BibitemShut
  {NoStop}%
\end{thebibliography}%

\end{document}